\documentclass[useAMS,usenatbib]{mn2e}
\usepackage{graphicx}
\usepackage{deluxetable}

\title[Newly Discovered CVs from IPHAS]{Newly Discovered Cataclysmic Variables from the INT/WFC Photometric H$\alpha$ Survey of the Northern Galactic Plane}

\author[A.R. Witham et al.]{
A. R. Witham,$^1$
C. Knigge,$^1$
A. Aungwerojwit,$^{2,3}$
J. E. Drew,$^4$
B. T. G\"ansicke,$^2$
\newauthor
R. Greimel,$^{5,6}$
P. J. Groot,$^7$
G. H. A. Roelofs,$^{7,8}$
D. Steeghs,$^{2,8}$
and P. A. Woudt$^9$
\\
$^{1}$ School of Physics \& Astronomy, University of Southampton, Highfield, SO17 1BJ, U.K.\\
$^{2}$ Department of Physics, University of Warwick, Coventry CV4 7AL, U.K. \\
$^{3}$ Department of Physics, Faculty of Science, Naresuan University, Phitsanulok, 65000, Thailand\\
$^{4}$ Imperial College of Science, Technology and Medicine, Blackett Laboratory, Exhibition Road, London, SW7 2AZ, U.K \\
$^{5}$ Isaac Newton Group of Telescopes, Apartado de correos 321, E-38700 Santa Cruz de la Palma, Tenerife, Spain \\
$^{6}$ Institute of Physics, University of Graz, Universitätsplatz 5, 8010 Graz, Austria\\
$^{7}$ Afdeling Sterrenkunde, Radboud Universiteit Nijmegen, Faculteit NWI, Postbus 9010, 6500 GL Nijmegen, the Netherlands \\
$^{8}$ Harvard-Smithsonian Center for Astrophysics, 60 Garden Street, Cambridge, MA 02138, USA \\
$^{9}$ Department of Astronomy, University of Cape Town, Rondebosch 7700, South Africa \\
}

\begin{document}

\date{2005 August 5}

\pagerange{\pageref{firstpage}--\pageref{lastpage}} \pubyear{2005}

\maketitle

\label{firstpage}

\begin{abstract}
We report the discovery of 11 new cataclysmic variable (CV) candidates by the Isaac Newton Telescope (INT) Photometric H$\alpha$ Survey of the northern Galactic plane (IPHAS). Three of the systems have been the subject of further follow-up observations.  For the CV candidates IPHAS J013031.90+622132.4 and IPHAS J051814.34+294113.2, time-resolved optical spectroscopy has been obtained and radial-velocity measurements of the H$\alpha$ emission-line have been used to estimate their orbital periods.  A third CV candidate (IPHAS J062746.41+ 014811.3) was observed photometrically and found to be eclipsing. All three systems have orbital periods above the CV period-gap of 2--3\,h. We also highlight one other system, IPHAS J025827.88+635234.9, whose spectrum distinguishes it as a likely high luminosity object with unusual C and N abundances.
\end{abstract}

\begin{keywords}
surveys -- binaries: close -- novae, cataclysmic variables
\end{keywords}

\section{Introduction}
\label{intro}
Cataclysmic variables (CVs) are semi-detached interacting binary systems
containing a white dwarf (WD) primary and a late-type main-sequence secondary.
The secondary fills its Roche lobe, and matter is transferred to the primary through the inner Lagrangian point. The mass-transfer process in CVs means that line emission is a common observational feature of the majority of CVs, especially in the Balmer series. Observed lines may originate from optically thin or irradiated parts of the accretion disc (if present, \citealt{1980ApJ...235..939W}). Line emission may also be produced in the accretion stream and on the irradiated face of the secondary star. \citet{1995cvs..book.....W} provides a comprehensive review of CVs.

Population synthesis models suggest that the intrinsically faintest low accretion rate systems should dominate the Galactic CV population and should be found predominantly at short orbital periods, that is, below the well-known CV ``period gap'' between 2 hrs and 3 hrs \citep{1993A&A...271..149K, 1997MNRAS.287..929H}. This dominant population of faint, short-period CVs has proven quite elusive. This is partly because most known CVs have been discovered with techniques
that are biased against detecting CVs with low mass 
accretion rates. For example, all flux-limited surveys with relatively
bright limiting magnitudes and also variability
searches (e.g. for dwarf nova outbursts) are intrinsically biased
against the detection of faint, short-period CVs with potentially long 
inter-outburst recurrence times (such as WZ Sge). For a closer
look at the period distributions of CVs found with different discovery
methods, see \citet{2005ASPC..330....3G}. Recently \citet{2007MNRAS.374.1495P} have shown that the relative dearth of short-period CVs is not just due to selection effects, implying a serious flaw in our understanding of CV evolution (see also \citealt{2006A&A...455..659A}).  Nevertheless, selection effects must be at least partly responsible for the lack of known faint short-period CVs, and so many such systems still remain to be found.


Given the ubiquity of line emission amongst CVs, searches for objects displaying H$\alpha$ emission
offer a powerful way to find new CVs. Examples of such
CV searches in the Galactic field include those of \citet{2002ASPC..261..190G}, \citet{2005A&A...443..995A}, \citet{2002A&A...395L..47H}, and the Chandra Multiwavelength Plane (ChaMPlane) survey (\citealt{2005ApJ...635..920G}, \citealt{2005ApJS..161..429Z} and \citealt{2006ApJS..163..160R}). Similar searches have also been performed in globular clusters have been performed by \citet{1995ApJ...439..695C}, \citet{1995ApJ...455L..47G}, \citet{1996ApJ...473L..31B} and \citet{2000ApJ...532..461C}. The Isaac Newton Telescope (INT) Photometric H$\alpha$ Survey of the northern Galactic plane (IPHAS) is currently surveying the Milky Way in broad-band Sloan $r^\prime$ and $i^\prime$ and narrow-band H$\alpha$ and provides an excellent data
base for a detailed CV search at low Galactic latitudes.  The survey
goes to a depth of $r^\prime\simeq20$\,mag and covers the latitude range
$-5^o < b < +5^o$. A detailed introduction to the survey, including the transmission profiles of the filters it uses, is given by \citet{2005MNRAS.362..753D}. What makes IPHAS particularly promising for finding CVs is that, empirically, the intrinsically  faintest, low mass transfer rate ($\dot{M}$) systems tend to have the largest Balmer line  equivalent widths (EWs; \citealt{1984ApJS...54..443P}; \citealt{2006MNRAS.369..581W}). Unless the inverse relationship between Balmer EW and orbital period breaks down for the faintest, short-period systems, emission-line surveys should be very good at finding low-$\dot{M}$ CVs. IPHAS should then be an excellent way of detecting a large population of faint, short-period CVs.

In this paper we announce the discovery of eleven new CV candidates from the IPHAS survey data. We present spectroscopic follow-up observations of two of these CVs candidates and determine their orbital periods. We also present the results of time-series photometry of another CV candidate which establishes it as a long period, eclipsing, system.

\section{Observations}
H$\alpha$ emitters were selected from ($r^\prime$ - H$\alpha$) versus ($r^\prime$ - $i^\prime$) colour-colour plots created using IPHAS photometry. Techniques for selecting H$\alpha$ emitters from such plots are described in \citet{2005MNRAS.362..753D}, \citet{2006MNRAS.369..581W} and Witham et al. (2007 MNRAS submitted). These techniques use fits to the upper locus of points in each colour-colour plot. Objects which lie significantly above the fits are selected as H$\alpha$ emitters. Whether an object is selected as an emitter thus depends on its ($r^\prime$ - H$\alpha$) colour, $r^\prime$ band magnitude and ($r^\prime$ - $i^\prime$) colour. The ($r^\prime$ - $i^\prime$) dependence is due to the upper locus (which generally tracks the unreddened main sequence) has a positive gradient and moves to larger values of ($r^\prime$ - H$\alpha$) as ($r^\prime$ - $i^\prime$) increases \citep{2005MNRAS.362..753D}. The $r^\prime$ band magnitude dependence is simply a result of the increasing noise at fainter magnitudes. Initial spectroscopic follow-up to IPHAS has already been used to identify hundreds of H$\alpha$ emitters, and this has so-far led to the discovery of 11 new CVs candidates. 

\begin{figure*}
\includegraphics[angle=0,width=0.8\textwidth]{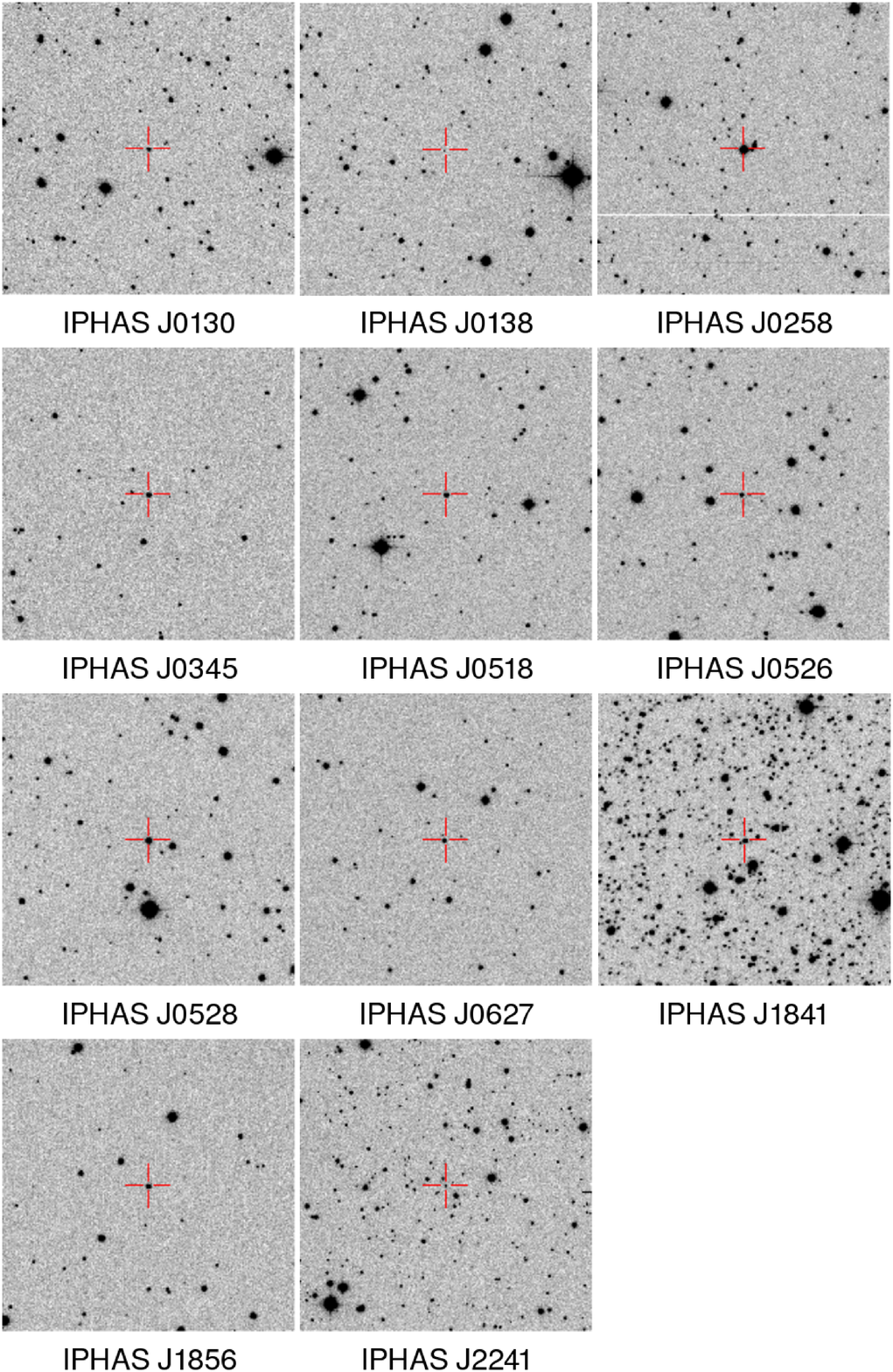}
\caption{\label{find_chart} IPHAS $r^\prime$ band finding charts showing the eleven new CV candidates identified by long-slit and Hectospec spectroscopy.  The charts are 3 by 3\,arcmin in size.  North is towards the top and East is to the left.}
\end{figure*}

Long-slit ID spectroscopy of H$\alpha$ emitters identified by IPHAS and brighter than $r^\prime=18$\,mag~has been performed on a variety of telescopes. This follow-up effort has led to the discovery of 8 out of our 11 new CV candidates in a sample of 582 objects. In addition to long-slit spectroscopy the IPHAS consortium is also carrying out a campaign of multi-object spectroscopic follow-up. The goal of this is to establish the location of different populations of objects on the IPHAS ($r^\prime-\rmn{H}\alpha$) versus ($r^\prime-i^\prime$) colour-colour plane. Spectroscopy of $\sim15000$ objects has been obtained, and the current analysis of this data has resulted in the discovery of a further three new CV candidates. Inspection of IPHAS ($r^\prime$ - H$\alpha$) versus ($r^\prime$ - $i^\prime$) colour-colour plots reveals that these three CVs are all obvious outliers from the main stellar locus and have a significant H$\alpha$ excess. IPHAS photometry of all 11 new CV candidates is given in Table~\ref{nu_cvs}.  Finding charts for the CV candidates are shown in Fig.\,\ref{find_chart}. These are taken from the $r^\prime$ band IPHAS observations.

Time resolved spectroscopy and photometry of IPHAS J0130, J0518 and J0627 has been performed to confirm their classification and obtain estimates of their orbital periods. Details of the observing runs are given in the following sections.

\subsection{Long-Slit ID Spectroscopy}
The Intermediate dispersion Spectrograph and Imaging System (ISIS, \citealt{1992iagv.book.....C}) on the 4.2\,m William Herschel Telescope (WHT) was used in service time to identify H$\alpha$ emitters using the following configuration.  The red arm of the spectrograph was equipped with the R316R grating, and the blue arm was equipped with the R300B grating.  This results in an observed wavelength range of 3200--5700\,\AA~in the blue and 5800--8150\,\AA~in the red at a resolution of $\sim3$\,\AA. 

The Calar Alto 2.2\,m telescope equipped with the Calar Alto Faint Object Spectrograph (CAFOS, \citealt{1998ugc.book.....M}) was used to observe H$\alpha$ excess objects in August 2004. The spectrograph was equipped with the G-200 grating which resulted in a wavelength coverage of 4250--8300\,\AA~at a resolution of $\sim10$\,\AA.

Another observing run took place in 2004, but this time the 2.56\,m~Nordic Optical Telescope (NOT) was used with the Andalucia Faint Object Spectrograph and Camera (ALFOSC)\footnote{http://www.not.iac.es/instruments/alfosc/} and grism 7. This 600\,gpm grism delivers a resolution of $\sim4$\,\AA, and, for these observations, a wavelength range of 3760--6825\,\AA.

The final long slit spectrograph used was the FAst Spectrograph For the Tillinghast Telescope (FAST, \citealt{1998PASP..110...79F}).  The 1.5\,m Tillinghast telescope is located at the Fred Whipple Observatory, Mount Hopkins, Arizona. For the ID spectroscopy, the spectrograph was equipped with the 300\,gpm grating.  The resulting resolution is $\sim3$\,\AA, and the wavelength coverage is 3480--7400\,\AA. The spectra were obtained in service time.

Identification spectra were obtained from the raw data using standard data reduction procedures and a variety of software packages.  Bias-subtraction, flat-fielding, and extraction with cosmic-ray rejection were followed by wavelength calibration. Correction for the instrumental response was only done for the CAFOS spectra, from observations of standard stars.

The CVs candidates identified by each telescope and the date of discovery are given in Table~\ref{nu_cvs}.

\subsection{Multi-Object Spectroscopy}
 The Hectospec Multifiber Spectrograph (see \citealt{2005PASP..117.1411F} and references within) is being used to perform the multi-object spectroscopy.  This 300-fiber spectrograph is attached to the 6.5\,m Multi Mirror Telescope (MMT).  For our observations, the 270\,gpm grating is installed in the spectrograph to give a wavelength coverage of 3700--9150\,\AA~at a resolution of $\sim5$\,\AA. The CVs candidates identified by Hectospec and the date of identification are given in Table~\ref{nu_cvs}.
 
 The spectra are extracted using the instrumental pipeline as described by Steeghs et al. (in preparation). Corrections for telluric absorption have been made, but flux calibration has not been performed.

 
\begin{table*}
\begin{minipage}{135mm}
\caption{\label{nu_cvs} IPHAS photometry and discovery details of the eleven new CV candidates. The astrometry has a precision of $< \sim 0.1$ arcsec with respect to 2MASS and the external accuracy of the photometry is $\sim0.1$\,mag.}
\setlength{\tabcolsep}{1.1ex}
\begin{tabular}{llllllll}
\hline Abbreviated\tablenotemark{a} & IPHAS name/position & $r^\prime$ & ($r^\prime - i^\prime$) & ($r^\prime - \rmn{H}\alpha$) & Discovery & Telescope & Instrument\\
Name & J[RA(2000)+Dec.(2000)] &&&&Date&&\\
&&&&&\scriptsize{YYMMDD}&&\\
\hline
IPHAS J0130 & J013031.89+622132.3 & 16.9 & 0.5 & 0.7 & 040810 & Calar Alto 2.2m & CAFOS\\ 
IPHAS J0138 & J013840.58+580811.2 & 19.8 & 0.5 & 1.3 & 051025 & MMT & Hectospec\\ 
IPHAS J0258 & J025827.88+635234.9 & 13.4 & 0.4 & 0.3 & 041219 & NOT & ALFOSC\\ 
IPHAS J0345 & J034511.59+533514.5 & 16.0 & 0.8 & 0.5 & 060203 & Tillinghast & FAST\\ 
IPHAS J0518 & J051814.33+294113.0 & 16.5 & 0.7 & 1.0 & 040201 & WHT & ISIS\\ 
IPHAS J0526 & J052659.00+291508.4 & 17.3 & 0.4 & 0.4 & 040201 & WHT & ISIS\\ 
IPHAS J0528 & J052832.69+283837.6 & 15.6 & 0.5 & 0.5 & 061114 & Tillinghast & FAST\\ 
IPHAS J0627 & J062746.39+014811.1 & 16.4 & 0.5 & 0.7 & 040202 & WHT & ISIS\\ 
IPHAS J1841 & J184127.11+053822.0 & 17.2 & 0.7 & 0.5 & 060605 & Tillinghast & FAST\\ 
IPHAS J1856 & J185647.14+070055.3 & 17.1 & 0.8 & 0.7 & 050512 & MMT & Hectospec\\ 
IPHAS J2241 & J224112.21+564419.0 & 18.5 & 0.5 & 0.6 & 050705 & MMT & Hectospec\\ 
\hline
\footnotetext[1]{This name is used only for the purposes of this paper for brevity and is not the full IPHAS name.}
\end{tabular}
\end{minipage}
\end{table*}



\subsection{Time-Resolved Spectroscopy}
\label{spect}
Time-resolved spectroscopy of IPHAS J0130 and IPHAS J0518 was obtained using CAFOS on the 2.2\,m telescope at the Calar Alto observatory.  The G-100 grism was used with the SITe-1d 2Kx2K CCD camera, giving a dispersion of 2.1\,\AA/pix and a wavelength range of 4250--8300\,\AA.  Two individual observing runs were allocated to the project, one in January 2005 and the second in November 2005.  In the January run, exposure times of 900\,s and a slit width of 1.2\,arcsec were used. In the November run, exposure times of 720\,s and a slit width of 1.2\,arcsec were used.  Mercury lamps and helium rubidium lamps were used to obtain arc spectra. Arc exposures were taken throughout each night, bracketing the exposures of target objects and typically being separated by at most one hour. Spectro-photometric standard stars were observed to correct for the instrumental response. Weather conditions and seeing were variable throughout both runs. Further details of the two observing runs are given in Table~\ref{tab1}.


A total of 38 exposures were obtained for J0130 and 27 exposures for J0518.  Data reduction and analysis were carried out using software written by Tom Marsh. Bias subtraction, flat-field correction, spectral tracking and optimal spectral extraction \citep{1989PASP..101.1032M} of the data were carried out using \textsc{pamela}.  Wavelength calibration and correction for instrumental response were done with \textsc{molly}. During wavelength calibration, the wavelength scale for each target frame was obtained by interpolating linearly between the two arc frames closest in time to the target frame.

\subsection{Time-Resolved Photometry}
\label{phot}
From November 30, 2004 to December 2, 2004, unfiltered time-resolved photometry of IPHAS J0627 was obtained using the University of Cape Town (UCT) CCD photometer \citep{1995BaltA...4..519O} on the 1.9\,m Telescope at the South African Astronomical Observatory (SAAO). Exposure times of 15, 20, and 30\,s were used, depending on the conditions and the target magnitude. Table~\ref{tab1} gives further details of the observations.

\begin{table}
\caption{\label{tab1} Details of the observing runs used to examine the temporal properties of three of the new CVs candidates from the IPHAS survey data.}
\begin{tabular}{llll}
\hline Start HJD & End HJD & Exp. &  Frames\\
&& (s) & \\
\hline
\textbf{J0130} & (CAFOS spectroscopy)\\
2453399.3355749 & 2453399.4093686 & 900 & 7\\
2453697.3989292 & 2453697.4597595 & 720 & 7\\
2453699.4175540 & 2453699.5389790 & 720 & 12\\
2453702.4216838 & 2453702.5349560 & 720 & 12\\
\\
\textbf{J0518} & (CAFOS spectroscopy)\\
2453699.5802004 & 2453699.7235635 & 720 & 15\\
2453702.5676267 & 2453702.6806411 & 720 & 12\\
\\
\textbf{J0627} & (SAAO photometry)\\
2453340.4204225 & 2453340.5314034 & 15 & 640\\
2453341.4031453 & 2453341.5372498 & 30 & 104\\
2453342.4685924 & 2453342.5652070 & 20/30 & 296\\
\hline
\end{tabular}
\end{table}

\section{Identification Spectra}
\begin{figure*}
\includegraphics[angle=0,width=\textwidth]{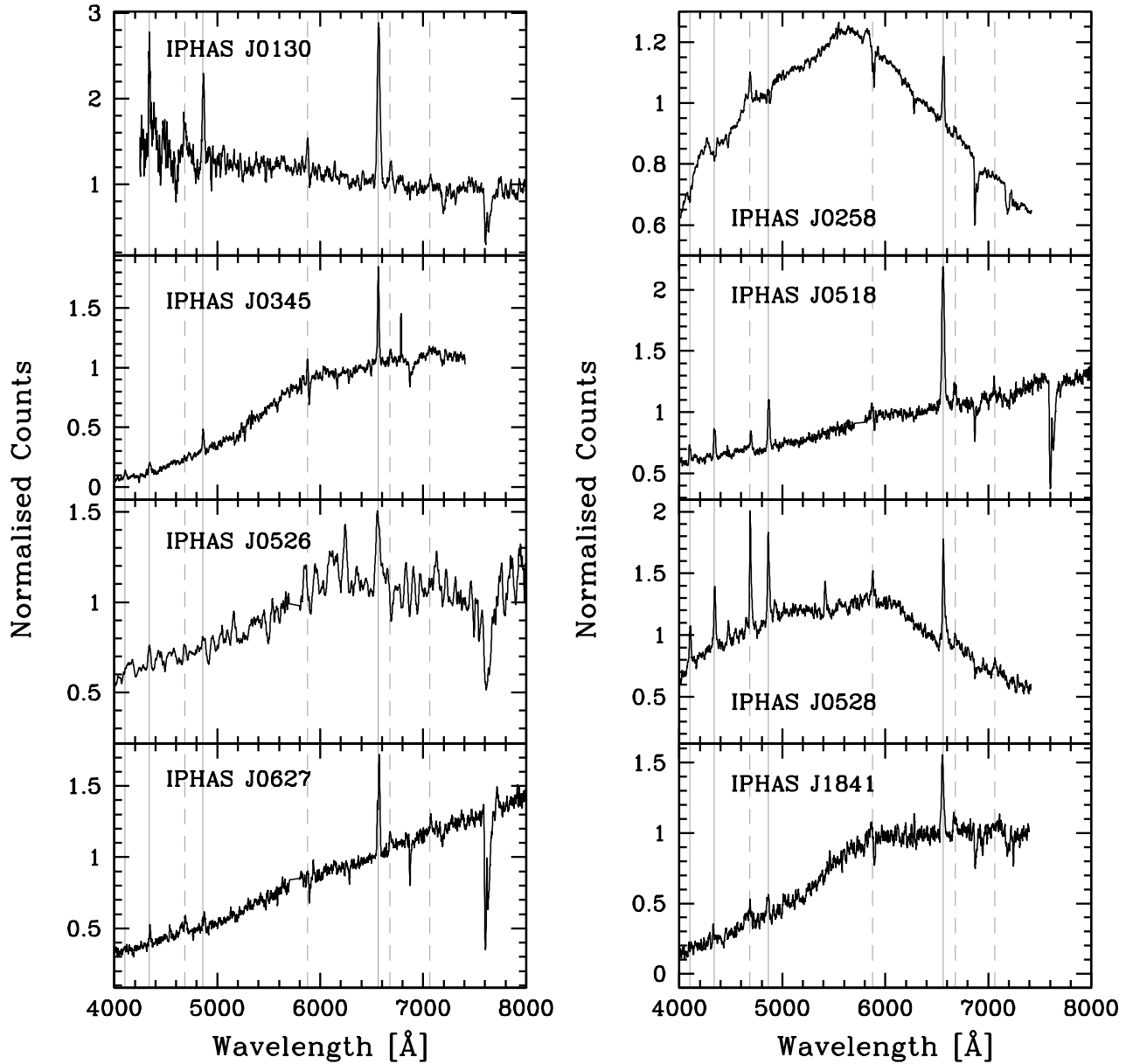}
\caption{\label{cv_id_spectra_bc} Identification spectra of the eight new CV candidates identified by long-slit spectroscopy. The spectra are not flux calibrated but have been normalised to the intensity of the continuum close to the H$\alpha$ emission-line. The spectrum of IPHAS J0130 has been flux calibrated but is shown normalised in the same fashion as the other spectra. Note that due to improved signal to noise the spectrum shown of IPHAS J0258 is from a Tillinghast/FAST observation not the earlier NOT/ALFOSC observation. Solid grey lines indicate the Balmer lines and dashed grey lines indicate the prominant He\,\textsc{I} and He\,\textsc{II} lines.  Note the presence of telluric absorption bands in the spectra, particularly those near 6900, 7200, and 7600\,\AA.}
\end{figure*}

\begin{figure}
\includegraphics[angle=0,width=\columnwidth]{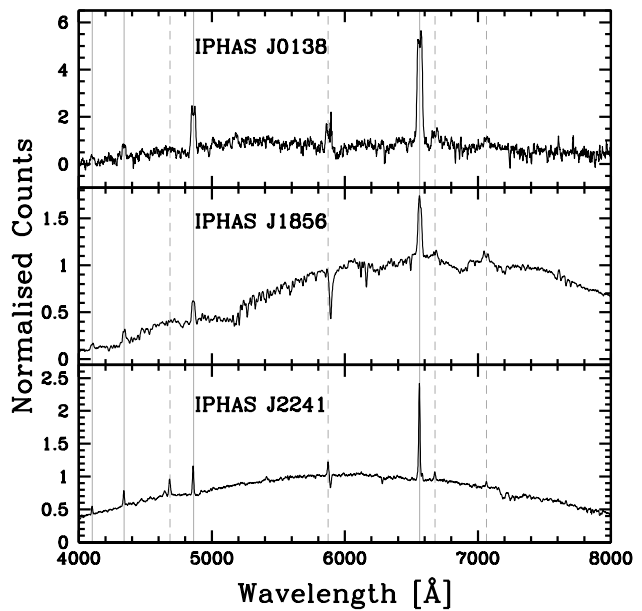}
\caption{\label{cv_id_spectra_hect} Identification spectra of the three new CV candidates identified by Hectospec.  The spectra are not flux calibrated and but been normalised to the intensity of the continuum close to the H$\alpha$ emission-line. Solid grey lines indicate the Balmer lines and dashed grey lines indicate the prominant He\,\textsc{I} and He\,\textsc{II} lines.}
\end{figure}

The identification spectra for the new CV candidates are presented in Fig.\,\ref{cv_id_spectra_bc} and Fig.\,\ref{cv_id_spectra_hect}. Other types of object which show H$\alpha$ emission and whose spectra can be similar to CVs include Be stars, active late-type stars and symbiotics. Symbiotics can be distinguished from CVs by the contribution of the absorption spectra of a late-type giant and the large EWs of the emission lines. Spectroscopic follow-up of IPHAS has already resulted in 4 new symbiotics being found (Corradi et al. 2007 A\&A in preparation). The median H$\alpha$ EWs of these systems and the already known symbiotics observed during IPHAS spectroscopic follow-up is $\sim400$\,\AA. Active late-type stars can be distinguished from CVs by the dominant late-type molecular continuum and the comparatively small EW of the H$\alpha$ emission-line, which is less than 10\,\AA~in these stars \citep{1989A&A...217..187P}. The Be stars observed by IPHAS are expected to be far more reddened than the CVs observed, as they are intrinsically much brighter objects. Also, in contrast to the composite continuum seen in CV spectra, the continuum of a Be star is purely that of an early-type star. Furthermore, Be stars in general have narrow emission lines compared to CVs. The full width at half maximum of the H$\alpha$ emission-line in Be stars is typically 100-500\,km$\rmn{s}^{-1}$ (see for example, \citealt{1982A&AS...48...93A}, \citealt{1983A&AS...53..319A}, \citealt{1986A&A...166..185H} and \citealt{2000A&AS..147..229B}), whereas in CVs with accretion disks line widths of several thousands of kilometres per second are possible \citep{1986MNRAS.218..761H}. However, not all CVs have broad lines, and it is possible that such CVs could be mistaken for Be stars, particularly if the absorption spectra of the donor is not apparent in the spectra.

The objects presented here have been classified as CV candidates by virtue of their broad Balmer and Helium emission lines.  Particularly helpful in classification is the broad He\,\textsc{I} 6678.2\,\AA~emission-line present in the spectra. This line is not expected to be present in the spectra of other types of system, and therefore identifies CVs effectively. The He\,\textsc{II} emission-line at 4685.7\,\AA~is also prominent in several of the spectra (for example IPHAS J0528).  The presence of strong He\,\textsc{II} emission lines, especially in comparison to the H$\beta$ line, is often a characteristic of magnetic CVs. IPHAS J0528 and J0258 are good examples of cases where the He\,\textsc{II} 4685.7\,\AA~line is much stronger than the H$\beta$ line.

The secondary star noticeably contributes to the spectrum in a few systems and lends further weight to the classification. IPHAS J1856, for example, appears to have a late K donor star by virtue of the broad molecular absorption bands present in the spectrum. Molecular bands present include those of the TiO molecule near 6200\,\AA, and 7200\,\AA. and the presence of the MgH triplet feature near $5200$\,\AA. 

Additional points worth noting from closer examination of the identification spectra are the presence of double-peaked Balmer lines in the spectra of IPHAS J0258, J0518, J0627 and J0138.  The double-peaks are indicative of the underlying accretion disc in a system which is observed at moderate to high inclinations.  Indeed, eclipses are observed in the SAAO photometry of J0627 which confirm that this system is viewed at high inclination.

Finally note that the signal-to-noise ratio in the spectrum of IPHAS J0526 is relatively low, and its classification as a CV candidate is tentative.  When compared to the other CV candidates observed using ISIS, the continuum of IPHAS J0526 appears much flatter toward the red end of the spectrum.

\section{H$\alpha$ EWs}
\begin{figure}
\includegraphics[angle=0,width=\columnwidth]{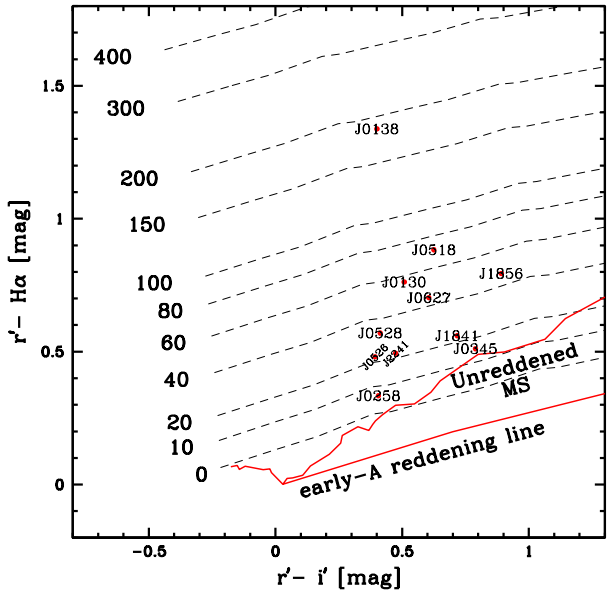}
\caption{\label{cv_col_col} IPHAS colour-colour plot showing the location of the 11 new CV candidates based on their average colours. The red dots mark the location of the CV candidates and the abbreviated name is used to indicate the corresponding CV. Red lines indicate the theoretical unreddened main sequence and the early-A reddening line and these are labelled appropriately.  The dashed lines indicate lines of constant H$\alpha$ EW. The corresponding values are labelled to the left of the lines.}
\end{figure}

As has been shown in \citet{2006MNRAS.369..581W} and \citet{2005MNRAS.362..753D}, it is possible to
estimate the H$\alpha$ EWs of CVs using the IPHAS
photometry and the tracks of constant H$\alpha$ EW obtained from synthetic photometry. A power-law spectral energy distribution (SED) appropriate to an optically thick accretion disc ($F_{\lambda} \propto \lambda^{-2.3}$) is assumed when carrying out the synthetic photometry. The IPHAS
colours of the CV candidates have been used to obtain average
H$\alpha$ EWs. To enable a visual comparison of the photometric colours and the
corresponding EWs of the 11 new CV candidates, we show in Fig.\,\ref{cv_col_col} the average colours of each CV candidate on an $(r^\prime - \rmn{H}\alpha)$ versus $(r^\prime - i^\prime)$
plot. The CV candidates are labelled by their abbreviated name. Red
lines indicate the location of the unreddened main sequence
and early - A star reddening line, and dashed lines indicate
lines of constant H$\alpha$ EW \citep{2005MNRAS.362..753D}. The distribution
of the new CV candidates in $(r^\prime - \rmn{H}\alpha)$ is consistent with
the distribution of the previously known CVs observed by IPHAS
to have an H$\alpha$ excess \citep{2006MNRAS.369..581W}.  Furthermore, the $(r^\prime - i^\prime)$ distribution of the new CVs falls in the most densely populated region of the distribution of known CVs.

\begin{figure}
\includegraphics[angle=0,width=\columnwidth]{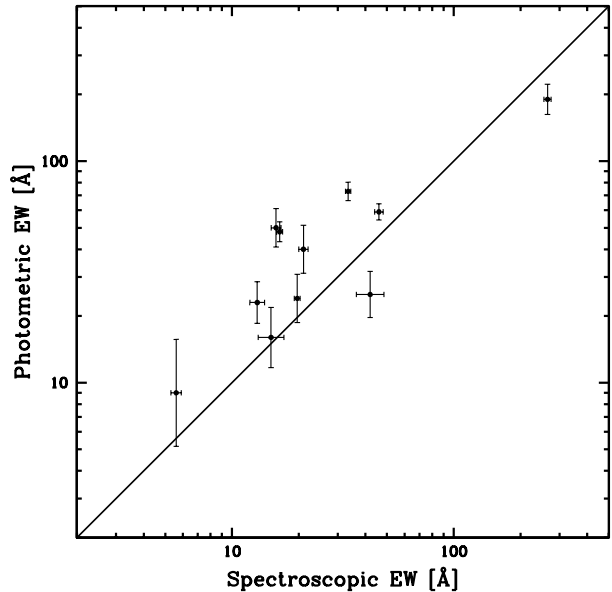}
\caption{\label{cv_ews}The correlation of the H$\alpha$ EWs of the 11 new CV candidates measured from the ID spectra and the IPHAS photometry. Spectroscopic EW uncertainties have been calculated using the techniques of \citet{1986MNRAS.222..809H}, however the systematic uncertainty in the continuum placement and the systematic error in the zero level has not been estimated. Photometric EW uncertainties have been estimated from the standard deviation of the EW values calculated from each available set of IPHAS photometry. A minimum uncertainty of 5\AA~is adopted because of the field-to-field scatter in the current IPHAS photometric calibration and because the power-law SED shape assumed may not be appropriate for all systems.}
\end{figure}

The H$\alpha$ EWs of the CV candidates measured from the ID spectra
have also been calculated for comparison with the photometric EWs. As shown in Fig.\,\ref{cv_ews}, the photometric and spectroscopic EWs correlate well, albeit with scatter and a bias toward higher photometric EWs. The scatter is not unexpected, because CVs can show significant H$\alpha$ EW variation over time. Indeed, even the fact that the EWs estimated from the photometry are generally greater than the spectroscopic EWs could be a selection effect: sources with variable H$\alpha$ EWs are more likely to be selected for follow-up if IPHAS observed them in a high-EW state.

\section{Additional Information on Selected Systems}
\subsection{IPHAS J0258}
This object distinguishes itself from the others whose spectra are shown in
Fig.\,\ref{cv_id_spectra_bc} and Fig.\,\ref{cv_id_spectra_hect} in possessing a broad emission feature at $\sim$5820\,\AA\ and clear absorption at 6284\,\AA\ (a well-known diffuse interstellar band [DIB]).
The emission is a strong hint that the abundance pattern is unusual.  This
possibility gains further support from the strength of 
He~{\sc ii}~$\lambda$4686\,\AA\ emission compared to the indistinct self-reversed
H$\beta$ feature, and the appearance of an emission complex at $\sim$4650\,\AA\ 
that includes N~{\sc iii}, C~{\sc iii} and C~{\sc iv} contributions.  Indeed, the broad feature at $\sim$5820\,\AA\ is also a complex blend attributable to these 
same ions, plus N~{\sc iv}.  The spectrum is also consistent with
the concept of Keplerian rotation in an accretion flow onto a compact object.  
The line emission in for example He~{\sc i}~$\lambda$6678\,\AA\ and at H$\beta$ is 
double-peaked with a peak-to-peak separation of $\sim$800~km~s$^{-1}$:  if interpreted
as due to Keplerian rotation around a solar mass object, the inferred 
orbit radius is $\sim$one solar radius or less.

The equivalent width of the DIB at 6284\,\AA\ is difficult to measure with any 
precision, because of an emission feature on its blue wing. A rough estimate 
would be 0.7\,\AA, which implies a minimum reddening of $A_V \sim 1.4$\,mag 
(scaling to the EW of this DIB in the reference object HD 183143; \citealt{1995ARA&A..33...19H}).  The maximum reddening likely for the sightline through the position 
of J0258 ($\ell = 136.37^\circ$, $b = 4.37^\circ$) is $A_V \simeq 3$\,mag \citep{1998ApJ...500..525S}.  The IPHAS and 2MASS photometry for this object indicate 
$A_V \sim 1.1$\,mag~and $A_V \sim 2.4$\,mag, respectively, if the dereddening is conducted on the 
basis that the intrinsic SED is that of an optically-thick accretion disk 
($F_{\lambda} \propto \lambda^{-2.3}$).   Adopting $1.5 < A_V < 2.0$\,mag, we 
estimate that the dereddened visual magnitude of J0258 is 11.5--12\,mag.  Being 
as reddened as it is, it is unlikely that J0258 is any closer than 
$\sim$1\,kpc (compare with distances and reddenings for open clusters within 5--10$^\circ$, listed in \citealt{2005A&A...438.1163K}).  If it is associated with the 
Perseus Arm, the distance estimate rises to $\sim$2\,kpc \citep{2006Sci...311...54X}.  Hence 
the likelihood is that this is a very luminous binary, with 
$M_V$ brighter than $\sim$2\,mag, and possibly as high as $\sim$0\,mag. 

The low contrast emission spectrum, the clues pointing toward enhanced C and 
N abundances, and the probable high intrinsic luminosity are all properties
reminiscent of the extreme systems V Sge \citep{1965ApJ...141..617H} and QU Car 
\citep{2003MNRAS.338..401D}.  J0258 is thus a newly uncovered member of an elite, but 
perplexing, group of interacting binaries.

\subsection{Radial-Velocity Studies of IPHAS J0130 and IPHAS J0518}
\label{res_s}
Measuring radial-velocity variations of the emission-line profiles in CVs is a useful technique for determining orbital periods and additional dynamical properties of the system. The large EWs of the H$\alpha$ line and its  location in an often flat part of the continuum make it an obvious candidate for measuring radial-velocity variations.

\begin{figure}
\includegraphics[angle=0,width=\columnwidth]{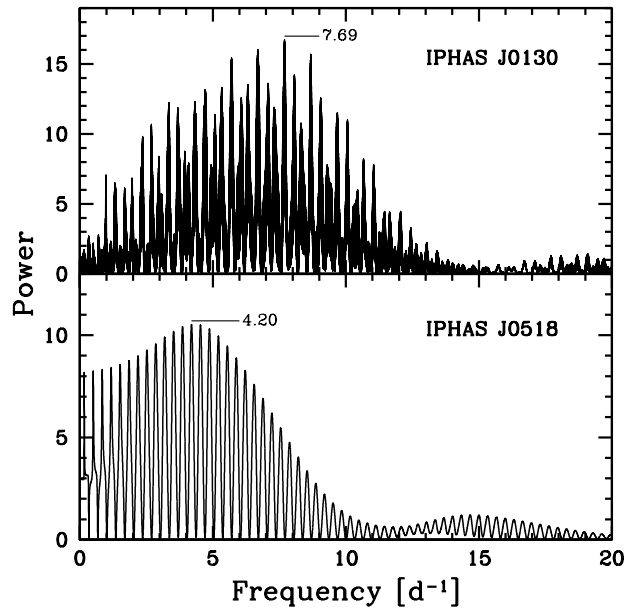}
\caption{\label{periodogram} The Lomb-Scargle periodograms obtained from the radial-velocity measurements.  Both periodograms are heavily aliased. The peak with the highest power is labelled in each case with the corresponding frequency.}
\end{figure}

Radial-velocity measurements of the data from IPHAS J0130 and J0518 were carried out using \textsc{molly}, once the data had been corrected to heliocentric Julian date (HJD), re-binned to a constant velocity scale, and normalised.  Single Gaussians were used to measure the position of the line core. However, double-peak structure and contamination from a bright spot can make accurate determinations of core radial-velocity variations impossible.  Therefore we also measured the radial-velocity variations using the double Gaussian technique of \citet{1980ApJ...238..946S}. A full width half maximum (FWHM) of 540\,$\rmn{km\,s}^{-1}$ was used for the Gaussians, which corresponds to one resolution element at H$\alpha$. The diagnostic method of \citet{1986ApJ...308..765S} was used to determine the best Gaussian separation for radial-velocity measurement. Time-series analysis of the resulting radial-velocity data was carried out using the Starlink programme \textsc{period}. The data was de-trended by subtracting the mean of
all the radial-velocity measurements from the combined nights and dividing by the standard deviation. Then Lomb-Scargle periodograms were created. The resulting periodograms, obtained from the radial-velocities measured, are shown in Fig.\,\ref{periodogram}.  There is a large amount of aliasing present in both periodograms, so it is not possible to accurately estimate the orbital period. \textsc{rvanal} was also used to give a better estimate of the orbital periods. A ``floating-mean'' periodogram \citep{1999ApJ...526..890C} is used by \textsc{rvanal} to analyse radial-velocity data. The ``floating-mean'' technique is similar to the Lomb-Scargle technique as it fits sinusoids to the data. However, unlike in the Lomb-Scargle method, the zero point of the sinusoid is a free parameter in the fits. This is an improvement, as it prevents power from being lost at long periods \citep{1999ApJ...526..890C}.




To further test the reliability of measured periods we used the Monte Carlo techniques of \citet{1985AJ.....90.2082T}. We used their definition of correctness-likelihood to evaluate the probability of a given peak in the periodogram being the orbital period.

\textsc{period} was used to fit a sine curve to the phase-folded radial-velocity data derived from the double Gaussian fit.  The parameters of the best fits are given in Table~\ref{tab2}. Using the phase zero measured by the sine fits and the estimates of the orbital periods, the spectra were binned into 10 phase-bins. The phase-binned spectra were then used to create trailed spectra.

\subsubsection{IPHAS J0130}
\label{radv_j0130}
\begin{figure*}
\includegraphics[angle=0,width=\textwidth]{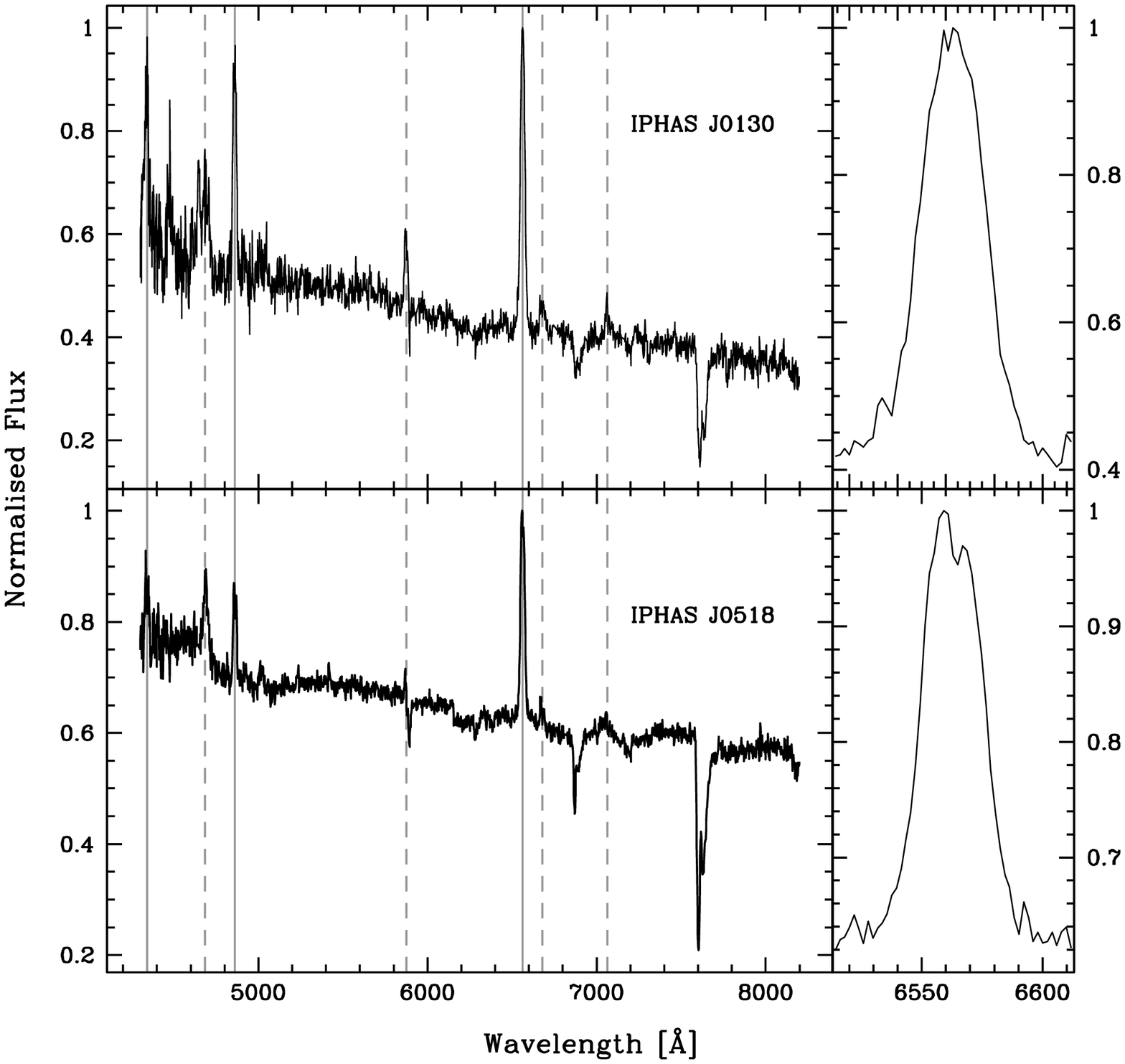}
\caption{\label{avg_spectra} The average spectra of IPHAS J0130 and IPHAS J0518 from the nights 27 (IPHAS J0130) and 24 (IPHAS J0518) November 2005. Solid grey lines indicate the Balmer lines and dashed grey lines indicate the prominant He\,\textsc{I} and He\,\textsc{II} lines. Note the clear Balmer emission and the presence of broad Helium emission lines characteristic of the spectra of CVs. Also note the presence of telluric absorption bands in the spectra, particularly those near 6900, 7200 and 7600\,\AA. The right hand panels in the figure are plots showing a close-up view of the H$\alpha$ emission-line.}
\end{figure*}

The averaged spectrum obtained on the night starting on November 27 is shown in Fig.\,\ref{avg_spectra}.  There is prominent H$\alpha$ emission, and the H$\beta$ emission-line is also present.  Helium emission lines can also be seen in the spectrum.  Due to the relatively low signal-to-noise, we only used the H$\alpha$ emission-line to measure radial-velocities.  In most of the individual spectra, the H$\alpha$ line appears single peaked, so a single Gaussian with a FWHM of 1400\,$\rmn{km\,s}^{-1}$ was used to measure the radial-velocity of the emission-line core.

The highest peak in the Lomb-Scargle periodogram of IPHAS J0130 corresponds to a period of 0.130062\,d. A value of 0.130061 $\pm$ 0.000001\,d ($\sim$3.12\,h) was obtained from \textsc{rvanal} (in agreement with the value obtained via the Lomb-Scargle periodogram), and we interpret this as the orbital period of the system. Results from the Monte-Carlo analysis show that the preferred period is indeed the strongest peak in the periodogram, and that this peak has a correctness-likelihood of 93 per cent.  The peak with the second highest power in the Lomb-Scargle periodogram has a longer period of 0.149209\,d ($\sim$3.58\,h), but has a correctness-likelihood of only 4 per cent.

\begin{figure}
\includegraphics[angle=0,width=\columnwidth]{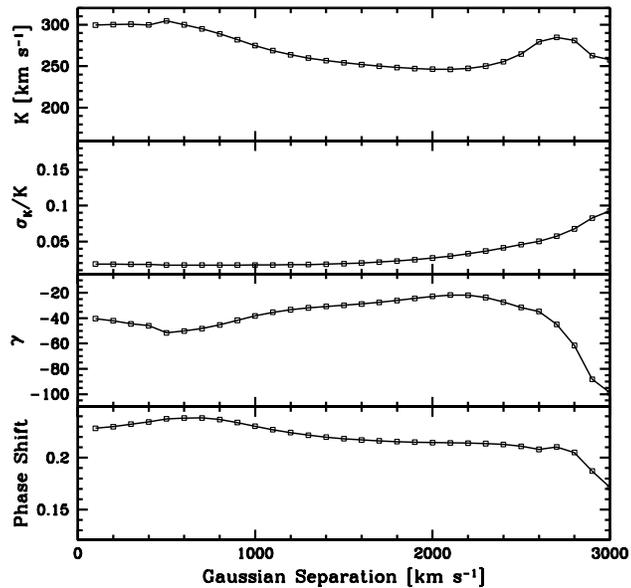}
\caption{\label{diagnostic_cv1} The radial-velocity diagnostic diagram for IPHAS J0130 showing the variations in phase shift, systemic velocity ($\gamma$), K--velocity and the fractional error in the K--velocity with the Gaussian separation used when measuring radial-velocities of the H$\alpha$ line. A FWHM of 540\,$\rmn{km\,s}^{-1}$ was used for the Gaussians. The phase shift represent the phase of the first radial-velocity data point in time relative to spectroscopic conjunction.}
\end{figure}

\begin{figure}
\includegraphics[angle=0,width=\columnwidth]{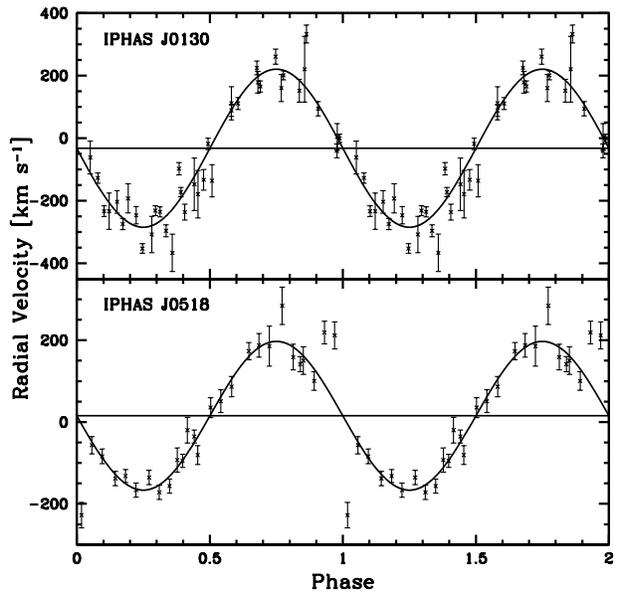}
\caption{\label{fold_sin} Sinusoidal fits to the radial-velocity data.  The data have been folded on the best-fit period and are repeated for clarity.}
\end{figure}

The diagnostic diagram obtained from measuring the radial-velocities with the double Gaussian method are shown in Fig.\,\ref{diagnostic_cv1}. The orbital period was fixed to be that obtained from the line-core.  Based on the diagnostic diagram, we used a value of 1600\,$\rmn{km\,s}^{-1}$ as the optimal Gaussian separation for further analysis. This value was chosen because at this separation the gradient of the line connecting the possible amplitude values is shallow, and the error in the amplitudes has not yet risen significantly. The radial-velocity data obtained using this Gaussian separation has been folded on the period estimate, and the results are shown in Fig.\,\ref{fold_sin} along with the sine curve fit to this data.
  
We have used the parameters of the sinusoid fit to the radial-velocity data to attempt to measure binary parameters of IPHAS J0130. Thus a secondary mass function was calculated under the assumption that the amplitude of the sinusoid represents the K-velocity of the primary. However, the secondary mass suggested by this analysis is much higher than expected for a CV at this orbital period ($M_2 \approx 0.20\,M_{\sun}$; \citealt{2006MNRAS.373..484K}) and is unlikely to be correct. Other reasonable choices for the Gaussian separations also resulted in unrealistically high mass functions. These results are not too surprising, considering the inconclusive nature of the diagnostic plot, which may be due to bright spot contamination.
 
\begin{figure}
\centering
\includegraphics[angle=0,width=\columnwidth]{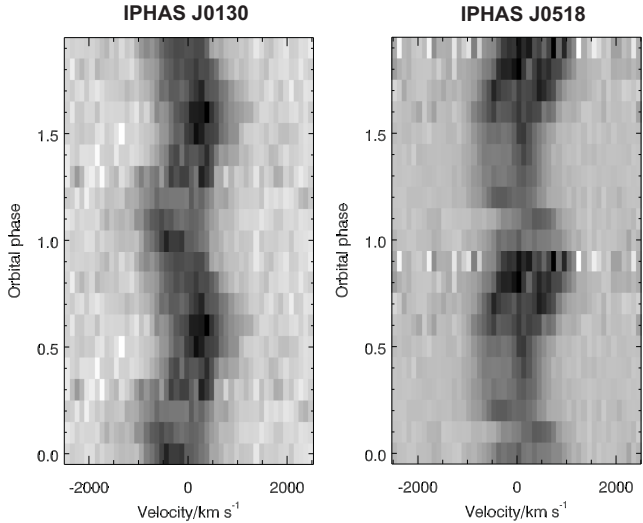}
\vspace{-0.3cm}
\caption{The phase-binned trailed spectra of IPHAS J0130 and IPHAS J0518. The spectra are centred on the rest wavelength of H$\alpha$.}
\label{fig:trail_spectra} 
\end{figure}

The phase-binned trailed spectra of IPHAS J0130 are shown in Fig.\,\ref{fig:trail_spectra}. The periodic nature of the motion of the H$\alpha$ line in the spectra is clearly apparent.





\subsubsection{IPHAS J0518}
\label{radv_j0518}
The average spectrum of J0518 from the night starting November 24 is shown in Fig.\,\ref{avg_spectra}. The spectrum shows strong H$\alpha$ and H$\beta$ emission.  The He\,\textsc{II} emission-line at 4686\,\AA~is also prominent.  Unfortunately, due to poor sky transmission on several of the later exposures, only the H$\alpha$ line could be used to measure the radial-velocity variations over the complete time-span of the observations. On closer inspection of the individual spectra around H$\alpha$, we see that the H$\alpha$ line is clearly separated into two peaks, due to the presence of an accretion disc in the system. As a result of the double-peaked structure of the H$\alpha$ line, the radial-velocity measurements of the core using a single Gaussian are unreliable.  Therefore, we have measured the radial-velocity variation using only the double Gaussian technique, with separations between 100 and 3000\,$\rmn{km\,s}^{-1}$.

\begin{figure}
\includegraphics[angle=0,width=\columnwidth]{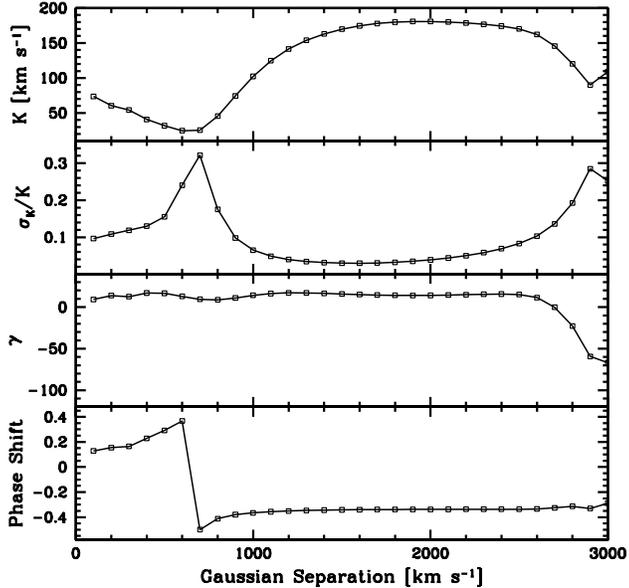}
\caption{\label{diagnostic_cv2} The radial-velocity diagnostic diagram for IPHAS J0518 showing the variations in phase, $\gamma$, K--velocity and the fractional error in the K--velocity with the Gaussian separation used when measuring radial-velocities of the H$\alpha$ line. A FWHM of 540\,$\rmn{km\,s}^{-1}$ was used for the Gaussians. The phases shown represent the phase of the first radial-velocity data point in time relative to spectroscopic conjunction.}
\end{figure}

A constant period of $\sim0.238$\,d was obtained from Lomb-Scargle periodograms of the de-trended data with separations from 900 to 2300\,$\rmn{km\,s}^{-1}$.  Sine curves with this period were used to fit the radial-velocity data and to construct the diagnostic diagram shown in Fig.\,\ref{diagnostic_cv2}.  At low values of the Gaussian separation, the Gaussians are located within the core of the profile, between the double peaks, so the results are unreliable. It is only once a separation of $\sim800$\,$\rmn{km\,s}^{-1}$ is exceeded that the wings are being sampled, and the results become reliable.  Conversely, at the highest velocities, the signal-to-noise is no longer high enough in the line wings. The radial-velocity data obtained with a Gaussian separation of 1900\,$\rmn{km\,s}^{-1}$ has been used for the remaining period analysis. The Lomb-Scargle periodogram obtained from this radial-velocity data is shown in Fig.\,\ref{periodogram} and has its highest peak at a period of 0.2382\,d. The refined period found using \textsc{rvanal} is 0.2383 $\pm$ 0.0007\,d ($\sim$5.72\,h). However, it is highly uncertain whether this is the true orbital period due to large power in the surrounding peaks. 


The parameters from the sinusoidal fit to the radial-velocity data with the period set to 0.2383 $\pm$ 0.0007\,d are given in Table~\ref{tab2}. The radial-velocity data were folded on the period found by \textsc{rvanal}, and the result is shown Fig.\,\ref{fold_sin}.  An attempt has been made to measure binary parameters of IPHAS J0158 in the same fashion as IPHAS J0130. In this case the diagnostic diagram looks more promising. Indeed, the expected secondary mass ($M_2 \approx 0.65\,M_{\sun}$, \citealt{2006MNRAS.373..484K}) is consistent with the measured mass function, provided the system is viewed at a reasonably high inclination. However, note that the long period of J0518 implies that the system may contain an evolved donor (\citealt{2006MNRAS.373..484K}).

It can be seen in Fig.\,\ref{fold_sin} that the radial-velocity values close to phase 0 are not fit well by the sinusoid, as the absolute values of these radial-velocities are much greater than the fits suggest.  This may be evidence of a rotational disturbance in the radial-velocity curve, caused by the secondary eclipsing parts of the accretion disc. The high inclination indicated by the rotational disturbance is consistent with the high inclination suggested by the secondary mass function.  The phase-binned trailed spectra of IPHAS J0518 shown in Fig.\,\ref{fig:trail_spectra} do not provide further evidence for a rotational disturbance, but the periodic nature of the motion of the H$\alpha$ line is clearly evident.



The results from the Monte-Carlo analysis show that four peaks have an appreciable correctness-likelihood: 0.2383\,d/$\sim5.72$\,h (26.4\%), 0.2203\,d/$\sim5.29$\,h (25.1\%), 0.2595\,d/$\sim6.23$\,h (20.6\%) and 0.2049\,d/$\sim4.92$\,h (10.7\%); the values in brackets are the correctness likelihoods expressed as percentages. The uncertainty on each of these peaks is 0.0007\,d. Even though we cannot decide between these four peaks with confidence, we can confirm that this CV lies above the period gap and is a long-period system.




\renewcommand\thefootnote{\thempfootnote}
\begin{table*}
\caption{\label{tab2} Parameters of the best sinusoid fits to the radial-velocity data. For comparison, we also show the fit parameters from the radial-velocity data from the core of the H$\alpha$ line for IPHAS J0130.}
\begin{tabular}{lllll}
\hline CV & Period & Amplitude & Phase zero\footnote{defined as the red-to-blue crossing point of the radial-velocities.} & $\gamma$ \\
& (d) & (\,$\rmn{km\,s}^{-1}$) & (HJD) & (\,$\rmn{km\,s}^{-1}$)\\
\hline
IPHAS J0130 (line--wings) & 0.130062 $\pm 1\times10^{-06}$ & 256 $\pm$ 13 & 2453550.756 $\pm$ 0.001 & -33 $\pm$ 10\\
IPHAS J0130 (line--core) & 0.130062 $\pm 1\times10^{-06}$ & 270 $\pm$ 11 & 2453550.757 $\pm$ 0.001 & -31 $\pm$ 9\\

IPHAS J0518 (line--wings) & 0.2383 $\pm$ 0.0007 & 187 $\pm$ 19 & 2453701.048 $\pm$ 0.004 & 18 $\pm$ 15\\
\hline
\end{tabular}
\end{table*}

\subsection{Photometric Variation in IPHAS J0627}
\label{photo_j0626}
\begin{figure}
\includegraphics[angle=0,width=\columnwidth]{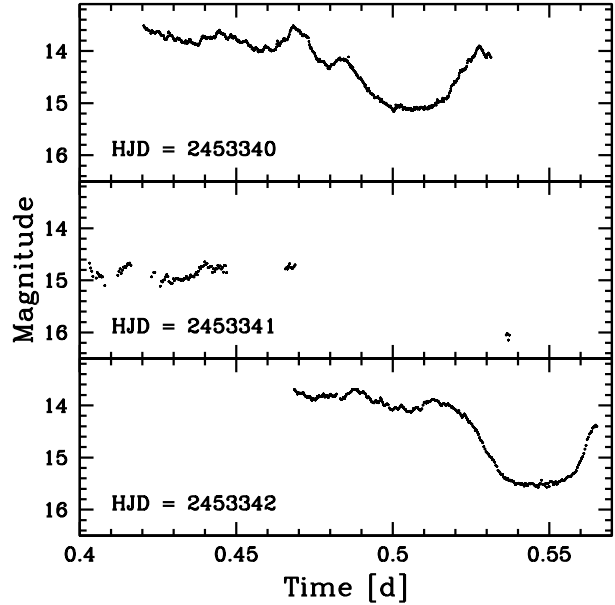}
\caption{\label{lightcurves} The lightcurves from the three nights of unfiltered photometry of IPHAS J0627 on the SAAO 1.9\,m telescope. The eclipses can clearly be seen on the first and third nights and the eclipse on the second night appears to have been caught by four data points.}
\end{figure}

 A total of 1040 individual exposures of IPHAS J0627 were obtained. Aperture photometry was used to calculate the relative magnitudes of the CV candidate. The light curves from each night of data obtained for IPHAS J0627 are shown in Fig.\,\ref{lightcurves}. Eclipses can be clearly seen in the light curves obtained from the first and third nights data.  In each case, the ingress of the eclipse has been observed, as well as the minimum light and parts of the egress. The light curve from the second night is sampled very sparsely, but an eclipse appears to have been caught by the final four data points obtained that night.
   
 

 To obtain an estimate of the lower limit on the orbital period, the duration of the total eclipse of the primary can be used. This eclipse duration is related to the orbital period and the inclination of the system due to the Roche geometry. For a mass ratio equal to 1, the ratio of the eclipse duration to the orbital period ranges from 0--0.13 as the inclination ranges from $68^\circ-90^\circ$ (\citealt{1976ApJ...208..512C}). Using this relationship it is possible to obtain a lower limit of $\sim0.25$\,d on the orbital period of J0627. This indicates that IPHAS J0627 is a long-period system.
 

In fact, it is possible to do better, because estimates of the orbital period can be obtained from the times of mid-eclipse. The eclipse is fairly well sampled in the data obtained on the first and third nights, but the photometry from the second night yields only a very rough eclipse timing. We thus estimate the orbital period by carrying out a weighted least squares fit to the 3 timings, allowing for their variable quality. Based on this, we find a set of 4 possible periods, $P_{orb} = 1.020 \pm 0.002$\,d, $0.5101 \pm 0.0008$\,d, $0.3401 \pm 0.0006$\,d and $0.2551 \pm 0.0004$\,d. Aliases at even shorter periods do exist, but we consider that these can be ruled out by the absence of multiple eclipses in individual runs, and the fact that the eclipse duration indicates a minimum orbital period of $\sim0.25$\,d. In choosing between these four possible periods the shape of the eclipse favours the shorter periods. More specifically, the flat bottom of the eclipses implies that the primary and the centre of the accretion disc are fully eclipsed, which indicates a high inclination.  This in turn means that higher ratios of the eclipse duration to the orbital period are favoured, and thus the shorter allowed periods are more likely.

\section{Discussion}
\label{discuss}
The first 11 CVs candidates presented here form the first batch of a potentially large sample of new CVs from the IPHAS survey data. Using the sample as it currently stands, it is possible to estimate the total number of bright, previously unknown CVs that IPHAS is expected to discover. So far 8 new CV candidates have been found with $r^\prime \leq 18$\,mag. These CVs candidates are from a sample of 582 bright ($r^\prime \leq 18$\,mag) IPHAS H$\alpha$ excess objects which have long-slit identification spectra. Therefore, the proportion of CVs in the total sample of H$\alpha$ emitters is 0.014. To the same limiting magnitude there are 3803 objects in a preliminary catalogue of H$\alpha$ emitters from IPHAS (Witham et al. submitted).  These have been obtained from 6142 IPHAS fields and their corresponding offset fields.  This gives a number density of 0.619 H$\alpha$ emitters per field at this limiting magnitude.  The total IPHAS survey will contain 7635 fields, so it is expected that IPHAS will uncover in the region of 4700 H$\alpha$ emitters (using the current selection techniques) down to this magnitude limit.  Therefore, using this simple extrapolation and assuming the proportion of CVs remains the same, $\sim65$ new bright CVs are expected to be found in the IPHAS survey.  This is a conservative number because the current sample of bright H$\alpha$ excess objects which have long-slit identification spectra are dominated by objects brighter than $\sim16.5$\,mag. As explained in the introduction, it is expected that the number of CVs should grow rapidly toward fainter magnitudes. Therefore, as the sample of objects becomes more complete to fainter magnitudes, the fraction of CVs found should increase. \citet{2006MNRAS.369..581W} have estimated that $\sim180$ short-period CVs could be detected by IPHAS

The results of the time-series analysis of three of the CV candidates in the sample have shown that all three lie above the 2--3\,h CV period gap.  This may seem surprising, as they were discovered due to their prominent H$\alpha$ emission.  However, \citet{2006MNRAS.369..581W} have shown that IPHAS is capable of detecting both long-period and short-period CVs as H$\alpha$ emitters, with the detection rate at fixed brightness having little dependence on the orbital period.  Combining this knowledge with the fact that short-period CVs are expected to be intrinsically fainter makes our detection of three long-period CVs less of a surprise, since all of these systems were selected for follow-up because they are bright. Bright CVs have also been discovered by virtue of their Balmer emission from the Hamburg Quasar Survey (HQS). These CVs are also predominantly long-period systems (see for example \citealt{2005A&A...443..995A}). However, we do expect to uncover new short-period CVs with IPHAS in the future, by observing fainter H$\alpha$ excess sources. Given that we only have 3 bright CVs candidates with period estimates in our sample so far, it is too early to make any tests of the population synthesis models.

\section{Conclusions}
 \label{conc}
Spectroscopy of several objects from the IPHAS H$\alpha$ excess catalogue has led to the discovery of 11 new CV candidates by virtue of their strong H$\alpha$ emission. The identification spectra of the CVs candidates include cases where the secondary star is particularly prominent (IPHAS J1856),
and examples of double-peaked emission lines (for example
IPHAS J0518). One particular candidate (IPHAS J0258) has similarities to the extreme objects V Sge and QU Car.

Time-resolved optical spectroscopy of two of the new CVs (IPHAS J0130 and IPHAS J0518) has been obtained using CAFOS on the 2.2\,m telescope at the Calar Alto Observatory. Radial-velocity measurements of the H$\alpha$ emission-line were used to determine their orbital periods.  Periodograms and a Monte Carlo analysis were used to estimate the orbital period of $P_{orb}=0.130061 \pm 0.000001$\,d for IPHAS J0130.  The periodogram obtained from the radial-velocity data of IPHAS J0518 reveals four peaks which could be the orbital period of the system: 0.2383, 0.2203, 0.2595 and 0.2049\,d, each with an uncertainty of 0.0007\,d. Time-series photometry of IPHAS J0627 using the SAAO 1.9\,m telescope found the system to be a long-period eclipser with four possible orbital periods: $1.020 \pm 0.002$\,d, $0.5101 \pm 0.0008$\,d, $0.3401 \pm 0.0006$\,d and $0.2551 \pm 0.0004$\,d. Further observations are necessary to obtain accurate values of the orbital period of all three systems, and to find accurate binary parameters.
\section*{Acknowledgments}
ARW was supported by a PPARC Studentship. AA thanks the Royal Thai Government for a studentship. BTG was supported by a PPARC Advanced Fellowship.  DS acknowledges a Smithsonian Astrophysical Observatory Clay Fellowship and a PPARC/STFC Advanced Fellowship. Tom Marsh is acknowledged for developing and sharing his software PAMELA, MOLLY, and RVANAL which were used for the reduction and analysis of spectroscopic data. Based in part on observations collected at the Centro Astron\'{o}mico Hispano Alem\'{a}n (CAHA) at Calar Alto, operated jointly by the Max-Planck Institut f\"{u}r Astronomie and the Instituto de Astrof\'{i}sica de Andaluc\'{i}a (CSIC).  This paper uses observations made at the South African Astronomical Observatory (SAAO).  This paper makes use of data from both the Isaac Newton and William Herschel Telescopes, operated on the island of La Palma by the ING in the Spanish Observatorio del Roque de los Muchachos of the Instituto de Astrofisica de Canarias.  Observations were obtained at the FLWO Observatory, a facility of the Smithsonian Institution. This research has made use of NASA's Astrophysics Data System.  This research has made use of the SIMBAD database, operated at CDS, Strasbourg, France. We acknowledge the data analysis facilities provided by the Starlink Project. Thanks to Luisa Morales Rueda for helpful comments on the data analysis. Thanks to Retha Pretorius for providing the code used to create the trailed spectra. Thanks to the WHT and MMT service observers. We thank the FAST queue observers at Fred Whipple Observatory for their
assistance, in particular Perry Berlind and Mike Calkins as well as the SAO
Telescope Data Center at the CfA for the data processing of FAST and
HectoSpec observations.
\bibliographystyle{mn_new}
\bibliography{mn-jour,arwbib}

\bsp

\label{lastpage}

\end{document}